\documentclass[12pt,eqsecnum,preprint]{aastex}
\begin{document}
\title{PKS~0743$-$67: An Ultra-luminous Accretion Disk and a High Kinetic Luminosity Jet}
\author{Brian Punsly\altaffilmark{1}}
\and\author{Steven Tingay\altaffilmark{2}} \altaffiltext{1}{4014
Emerald Street No.116, Torrance CA, USA 90503 and International
Center for Relativistic Astrophysics, I.C.R.A.,University of Rome
La Sapienza, I-00185 Roma, Italy, brian.m.punsly@L-3com.com or
brian.punsly@gte.net} \altaffiltext{2}{Centre for Astrophysics and
Supercomputing, Swinburne
  University of Technology, P.O. Box 218, Hawthorn, Vic 3122, Australia, stingay@astro.swin.edu.au}
\begin{abstract}
In this letter, deep radio observations of the quasar
PKS~0743$-$67 are presented that reveal a central engine capable
of driving jets with enormous kinetic luminosity, $Q>4.1 \times
10^{46}\mathrm{ergs/s}$. This result is significant because
archival optical spectral data indicates that the accretion disk
has a thermal luminosity, $L_{bol}>2\times
10^{47}\mathrm{ergs/s}$. Furthermore, estimates of the central
black hole mass from line widths indicate that
$L_{bol}/L_{Edd}\approx 1$. This suggests that neither a large
$L_{bol}$ nor $L_{bol}/L_{Edd}$ suppresses jet power in quasars,
despite claims that they do in the recent literature. Earlier
studies have found $L_{bol}$ and $Q$ are correlated in blazars.
However, by removing the BL-Lacs and leaving only the quasars in
the sample, we found that $Q$ is very weakly correlated with
$L_{bol}$ in the subsample.
\end{abstract}

\keywords{quasars: general --- quasars: individual (PKS~0743$-$67)
--- galaxies: jets --- galaxies: active --- accretion disks --- black holes}

\section{Introduction}The connection between accretion flow
parameters and radio jet production is a mysterious one. It has
been argued in \citet{wan04} that the jet kinetic luminosity, $Q$,
is correlated with the bolometric luminosity of the thermal
emission, $L_{bol}$, produced by the accretion flow in blazar type
AGN. However, using virtually identical techniques as
\citet{wan04}, \citet{cel97} came to the opposite conclusion. In
order to shed some light on this issue, we explore this question
from a different perspective for the particular case of quasars.
The vast majority ($\sim 90\%$) of quasars are radio quiet whether
their $L_{bol}$ lies just above the Seyfert 1/ quasar dividing
line or if they are at the other extreme,
$L_{bol}>10^{47}\mathrm{ergs/s}$. This observation indicates that
there are additional parameters, beyond $L_{bol}$ that
$L_{bol}/L_{Edd}$, that control the power of the radio jet. We
note that the very high $Q$, FRII radio source, Cygnus A, $Q
\approx 1.6\times 10^{46}\mathrm{erg/s}$ (according to (1.1) of
this article), harbors a hidden quasar with $L_{bol}$ just above
the Seyfert 1/ quasar dividing line and has a low Eddington rate,
the ratio of $L_{bol}$ to the Eddington accretion rate,
$L_{bol}/L_{Edd}\sim 0.01$ \citet{smi02,tad03}. Cygnus A is an
extremely powerful FR II radio source even when compared with low
frequency selected samples at high redshift \citep{wil99}. It has
two orders of magnitude higher $Q$ than most FR II quasars (see
Chapter 10 of \citet{pun01} and references therein). Thus, Cygnus
A provides a well studied "standard" candle for an extremely
powerful FR II source. This motivated us to explore the opposite
extreme in the quasar family, the very powerful quasar
PKS~0743$-$67, which is luminous in all frequency bands and seemed
to be a likely candidate for extremely high $Q$ jets. In sections
3 and 4, it is shown that it has a has a powerful accretion
luminosity, $L_{bol}> 2\times 10^{47} \mathrm{ergs/sec}$,
$L_{bol}/L_{Edd}\approx 1$, a strong unresolved VLBI radio core
and prominent radio lobes. Even though the quasar is at a redshift
of $z=1.511$ \citep{bec02}, both the radio core and the radio lobe
flux densities are $\sim 1Jy$.
\par In section 5, it is demonstrated that the high $Q$ for these two extreme ends of the quasar
range, Cygnus A and PKS 0743-067, are not out of line with the
properties of the quasar population as a whole. By studying a
sample of quasars from \citet{wan04}, we find that $Q$ is not
correlated with $L_{bol}$ for radio loud quasars that possess
blazar cores. Secondly, we demonstrate that the inverse
correlation claimed between $Q/L_{bol}$ and $L_{bol}/L_{Edd}$ in
\citet{wan04}, although true, is a trivial consequence of the fact
that $Q$ is not correlated with $L_{bol}$ in quasars. The primary
conclusion of this study is that the intrinsic power of a quasar
jet is not, to first order, controlled by the accretion rate.
\par We
have performed deep radio observations with the Australia
Telescope National Facility (ATCA) in order to understand the
radio structure of PKS~0743$-$673; the lobe emission alone would
qualify it for the 3C catalogue if the source were in the Northern
Hemisphere, our observations indicate that the jet kinetic
luminosity, $Q$, is far more powerful than that in Cygnus A.
\section{The Radio Observations}Previously, \citet{ray02} imaged the radio structure
of PKS~0743$-$673 at 4.8 GHz with the ATCA.  We performed deep
observations at 2.496, 4.800 and 8.640 GHz in order to image the
source structure and obtain higher resolution as well as spectral
and polarization data. It is essential to obtain both higher
resolution and accurate spectral data to assess the energy content
of the extended structure.  Our 8.640 GHz map is shown in Figure
1.
\begin{figure}
    \begin{center}
        \includegraphics[scale=0.8]{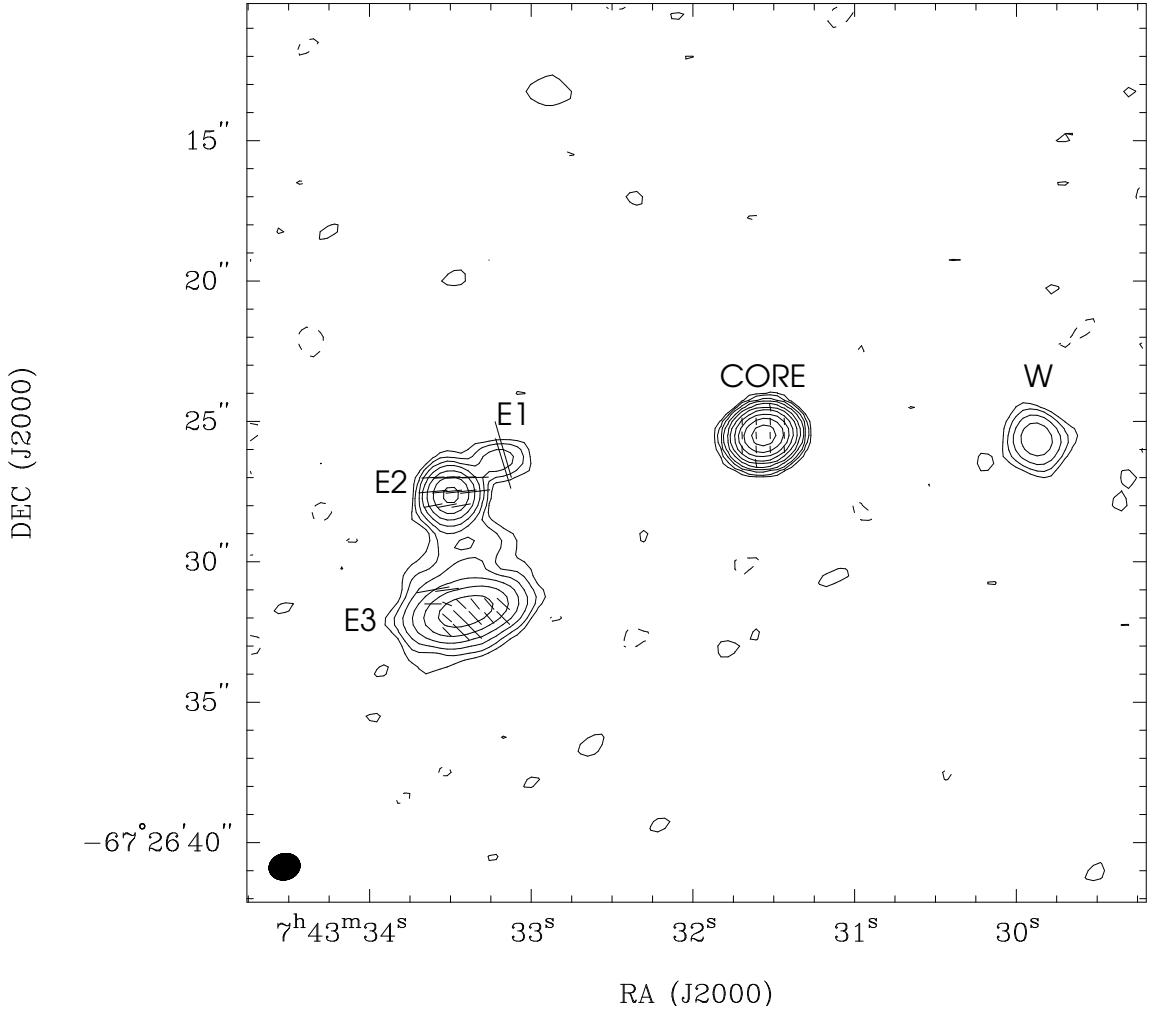}
    \end{center}
    \caption{PKS 0743-673 at 8640 MHz.  The peak intensity in the image is 1.25 Jy/beam.
    The beam-size is 0.9"$\times$1.1" at a position angle of $-78.8^{\circ}$.
    Contour levels for the Stokes I emission are 0.0125 Jy/beam $\times$
    ($-0.125$, 0.125, 0.25, 0.5, 1, 2, 4, 8, 16, 32, 64).  The peak fractional polarization
    is 52.1\%.  The vector lengths represent the electric field with 25.4\% fractional polarization per arcsecond.
    The beam ellipse is plotted in the lower left hand corner of the figure}
    \end{figure}

The data from our ATCA observations are presented in Table 1. Quasars with a strong flat spectrum core often have Doppler
enhanced kpc scale jets \citep{pun95}. Thus, an estimate of
$Q$ in PKS~0743$-$67 requires an analysis of the data in
Table 1 in order to determine if the jet and lobe emission to
the east of the nucleus is Doppler enhanced or not. The components E1, E2, E3
denote the eastern jet/lobe components in Figure 1, numbered consecutively
from west to east.

\begin{table}
\begin{center}
\begin{tabular}{ccccc} \hline \hline
$\nu$ &   Comp.& $S$     & $m$    & $\alpha$ \\
   (GHz)    &           &   (Jy)     & \%   & \\ \hline \hline
2.496      &  W      &  0.17    & 11       &...\\
          &   C      &  1.34    & 7        &...\\
          &   E1    &  ...    & ...     &...\\
          &   E2    &  0.53    & 11       &...\\
          &  E3    &  0.75    & 14       &...\\ \hline
4.800      &  W     &  0.08    & ...     &1.16\\
          &  C     &  1.17    & 8        &0.21\\
          &  E1   &  0.02    & 28       &...\\
          &  E2   &  0.31    & 22       &0.82\\
          &   E3   &  0.45    & 15       &0.78\\
 \hline
8.640      &  W     &  0.04    & ...     &1.17\\
          &  C     &  1.25    & 5        &0.06\\
          &  E1   &  0.01    & 52       &...\\
          &   E2   &  0.16    & 26       &0.98\\
          &   E3   &  0.24    & 15       &0.93\\ \hline \hline
\end{tabular}\\
\caption{ATCA radio data for PKS~0743$-$673.  Column 1: $\nu$, radio frequency.  Column 2:
 Component identification from Figure 1.  Column 3: $S$, total flux density at $\nu$.
 Column 4: $m$, percentage polarization at $\nu$.
Column 5: $\alpha$, two point spectral index between $\nu$ and 2.496 GHz
($S\propto\nu^{-\alpha}$).}
\end{center}
\end{table}
In Figure 1, the magnetic field (perpendicular to the electric
field vectors plotted) at the core is parallel to the jet
direction and remains parallel to the jet direction along the
length of the jet, even though the eastern jet goes through a
large apparent bend.  At the end of the eastern jet, the magnetic
field switches to being perpendicular to the jet direction,
typical of a radio galaxy hot spot.
\section{Estimating the Jet Kinetic Luminosity}
In order to avoid the ambiguities associated with Doppler
enhancement, we estimate the jet kinetic luminosity from the
isotropic extended emission, applying a method that allows one to
convert 151 MHz flux densities, $F_{151}$, measured in Jy, into
estimates of kinetic luminosity, $Q$, from \citet{wil99,blu00} by
means of the formula derived in \citet{pun05}:
\begin{eqnarray}
&& Q \approx 1.1\times
10^{45}\left[(1+z)^{1+\alpha}Z^{2}F_{151}\right]^{\frac{6}{7}}\mathrm{ergs/sec}\;,\\
&& Z \equiv 3.31-(3.65) \nonumber \\
&&\times\left(\left[(1+z)^{4}-0.203(1+z)^{3}+0.749(1+z)^{2}
+0.444(1+z)+0.205\right]^{-0.125}\right)\;,
\end{eqnarray}
where $F_{151}$ is the total optically thin flux density from the
lobes (i.e., no contribution from Doppler boosted jets or radio
cores). We assume a cosmology with $H_{0}$=70 km/s/Mpc,
$\Omega_{\Lambda}=0.7$ and $\Omega_{m}=0.3$.  In order to
implement this technique, one needs to determine which components
are optically thin and which are Doppler enhanced.

There are two possible interpretations of the data that one can
use to calculate $Q$. The most straightforward approach is to note
that all of the emission is optically thin and the large angular
size of the source, $\approx 250\,\mathrm{kpc}$, argues against
significant Doppler enhancement of the large-scale structures.
However, we choose the most conservative approach: assume that all
of the eastern emission is part of a jetted system and it is all
Doppler enhanced, even the hot spot to some extent (this would
explain why the eastern hot spot is more luminous than the western
hot spot). If the source were symmetric and viewed in the sky
plane then an upper limit to the total flux would be twice the
observed flux from the western hotspot, 340 mJy at 2.496 GHz.
Extrapolating this to 151 MHz, yields a lobe flux of 8.8 Jy.
Inserting this value into (3.1) yields $Q= 4.1 \times 10^{46}$
ergs/sec. This equates to 2.5 times the kinetic luminosity of
Cygnus A computed by the same method. If the eastern lobe is not
Doppler enhanced then the kinetic luminosity is even larger. We
note that no 151 MHz observations of PKS~0743$-$673 have been
made.  However, \citet{lar81} measured the total flux density at
408 MHz to be 8.6 Jy.  This measurement will be dominated by the
extended emission of the source, making an estimate of 8.8 Jy at
151 MHz for the unbeamed emission conservative.
\par A 2.3 GHz VLBI measurement of PKS 0743-67 was made in
\citet{pre89}. A secondary unresolved radio structure, presumably
a strong knot in a jet, is directed to the east of the core
towards the base of the kpc jet seen in figure 1. The VLBI
emission is dominated by an unresolved core on the 10
milliarcsecond scale with 1.2 Jy. Not only is the time averaged
$Q$ from PKS 0743-67 enormous, but the powerful parsec scale core
indicates that the source is still likely to be highly energetic
at the current time.
\section{Estimating the Eddington Ratio}One can estimate $L_{bol}$ as in
      \citet{lao98}, $L_{bol}\approx 8.3\nu L_{\nu}(3000\AA)$, a method that has
been applied to both radio quiet and radio loud quasars. We apply
this formula to the flux density at $3000\AA$ from the spectrum of
PKS 0743-67 in \citet{ali94}, yielding $L_{bol}\approx 4.7\times
10^{47} \mathrm{ergs/s}$. When making an estimate of the accretion
flow luminosity, the strong radio core might raise some concern
about contamination of the optical emission via a high frequency
synchrotron spectrum associated with the base of the jet. Thus,
alternatively, one could get an estimate of $L_{bol}$ using the
method of \citet{wan04} that depends on line luminosity. Following
the discussion in section 3 of \cite{wan04}, the CIV/Ly$\alpha$
line strength ratio of the composite quasar spectra in
\citet{fra91}, and eqn(1) of \citet{wan04} implies that the total
broad line luminosity is $L_{BLR}=8.83L_{CIV}$, where $L_{CIV}$ is
the CIV line strength. Secondly, \citet{wan04} estimate
$L_{bol}\approx 10 L_{BLR}$, thus $L_{bol}\approx 88.3 L_{CIV}$.
Using the CIV line strength from \citet{ali94} this implies
$L_{bol}\approx 2.91\times 10^{47} \mathrm{ergs/sec}$, in close
agreement with the estimate above from the spectrum directly.
\par One can estimate $L_{bol}/L_{Edd}$ using the $L_{bol}$
value above in conjunction with a mass estimate of the black hole
mass, $M_{bh}$, from the same CIV emission line. The estimator of
$M_{bh}$ of \citet{ves02} requires the luminosity at $1350\AA$,
$\lambda L_{\lambda}(1350\AA)$. To be consistent with the
philosophy of not using the continuum spectrum, one can instead
estimate  $\lambda L_{\lambda}(1350\AA)$ from the $L_{bol}$ that
is derived from the CIV line strength above with the aid of the
relation from \citet{lao98}, $L_{bol}\approx 8.3\nu
L_{\nu}(3000\AA)$ and assuming a typical quasar optical spectral
index of 0.7 as was done in \citet{wan04} (the spectrum in
\citet{ali94} yields a similar value, 0.75). One finds a central
black hole mass of $M_{bh}=1.62\times 10^{9} M_{\odot}$ and
$L_{bol}/L_{Edd}=0.99$. One can check this result independently
using the H$\alpha$ line of PKS 0743-67 measured in \citet{esp89}
and the estimators in \citet{gre05}, $M_{bh}=1.41\times 10^{9}
M_{\odot}$ and $\lambda L_{\lambda}(5100\AA)= 2.27\times
10^{46}\mathrm{ergs/s}$. Converting the line luminosity to
$L_{bol}$ as for the CIV estimate above, one finds $L_{bol}=
2.21\times 10^{47}\mathrm{ergs/s}$ and $L_{bol}/L_{Edd}=0.87$.
\begin{figure}
\epsscale{1.05}
\plottwo{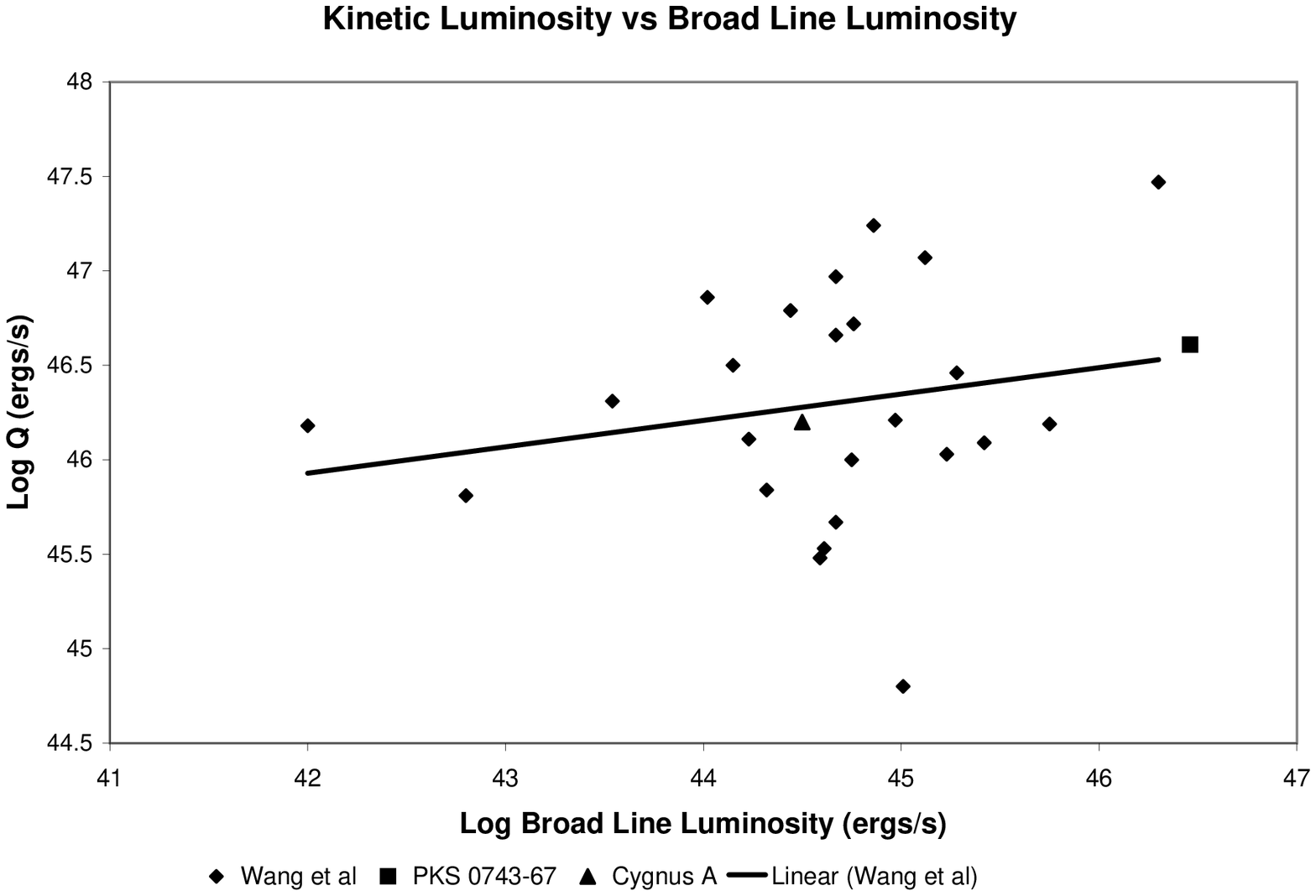}{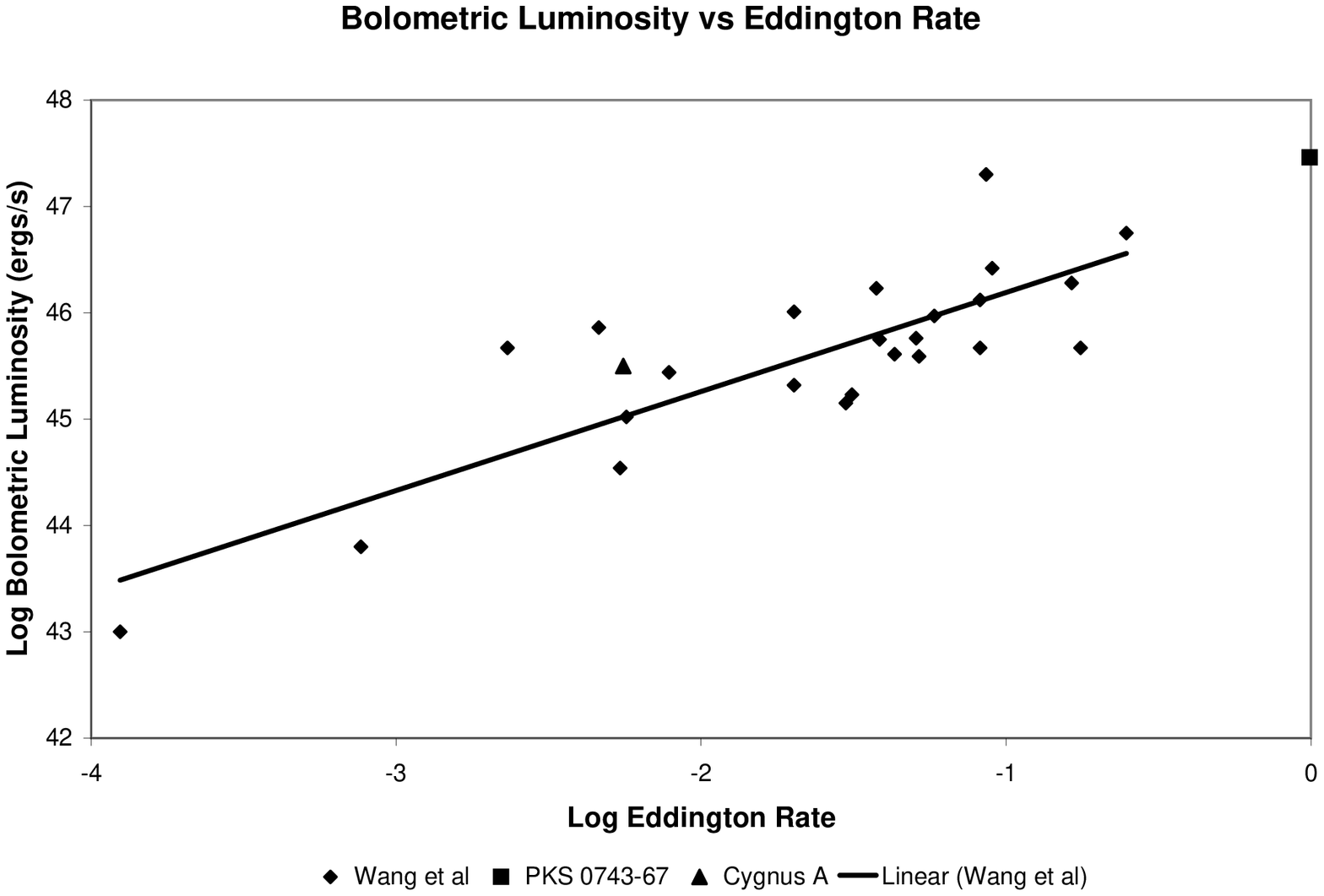}\\
\epsscale{1.10} \plottwo{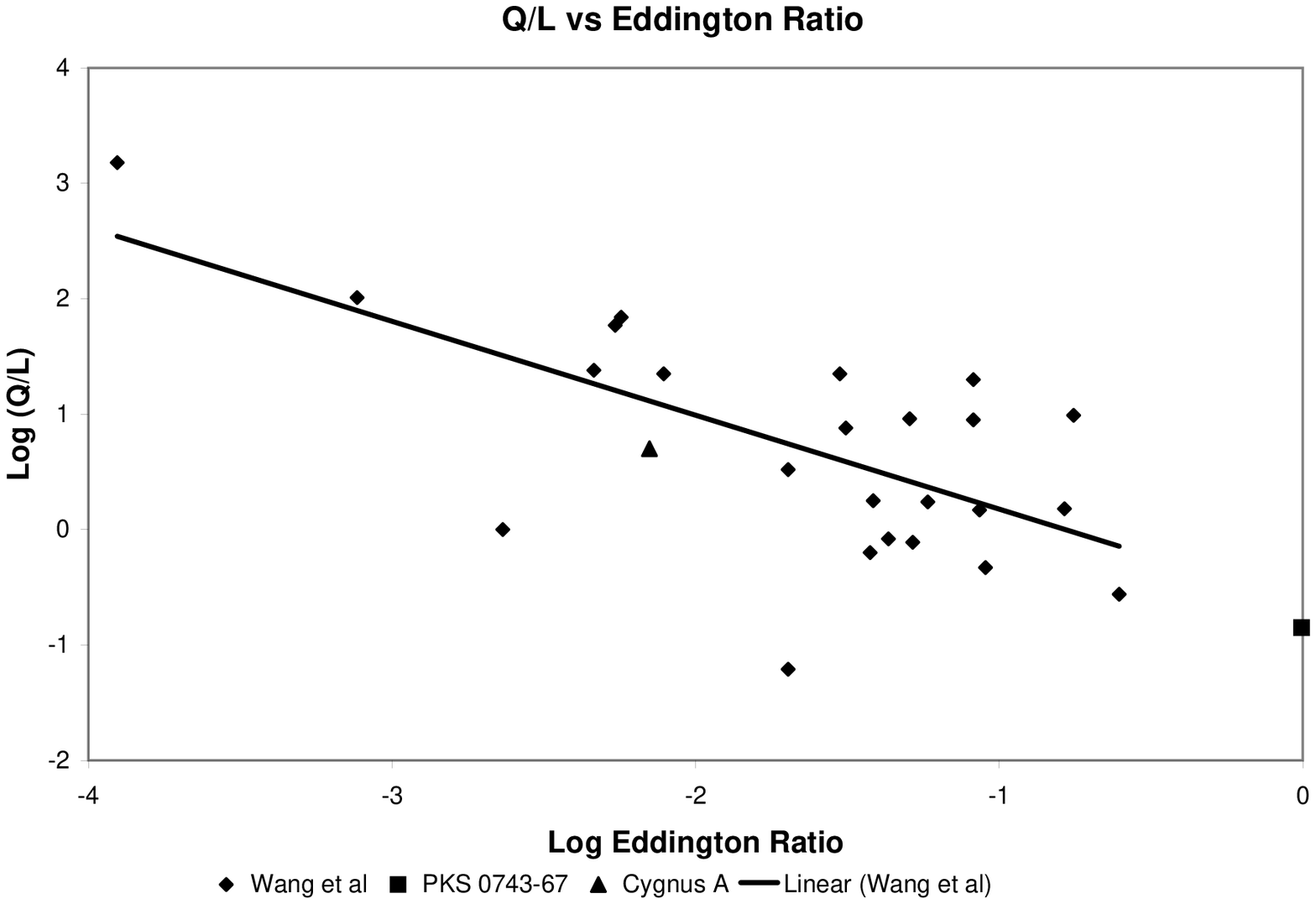}{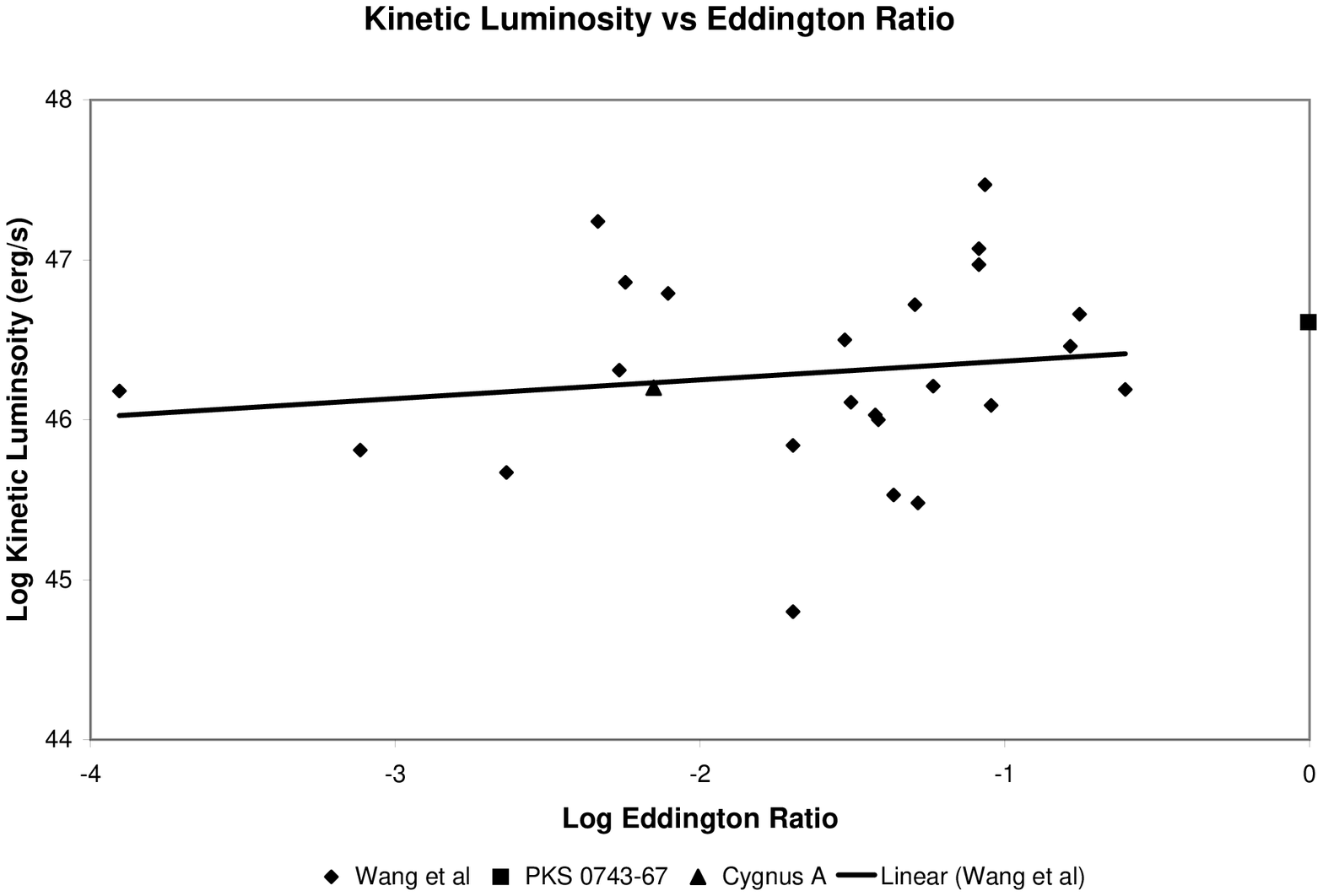}\caption{Figure 2a
(upper left hand corner) is a scatter plot of the logarithm of the
kinetic luminosity, $Q$, of the radio jets of in the quasar
subsample of \cite{wan04} versus the logarithm of the line
luminosity which they use as a measure of $L_{bol}$.
  The extreme estimate for $Q$ of 4C 52.27 noted in the text is omitted in this plot.
  The correlation is very weak. The estimates of $Q$ for Cygnus A and PKS
  0743-67 are added for sake of comparison and are based upon the isotropic
  methods of this paper, unlike the \citet{wan04} sample. A best fit
  line to the \citet{wan04} data is indicated. Figure 2b (upper right hand corner)
  is a scatter plot of the
logarithm of $L_{bol}$ versus the logarithm of $L_{bol}/L_{Edd}$.
Figure 2c (lower left hand corner) is a scatter plot of the
logarithm of $Q/L_{bol}$ versus the logarithm of
$L_{bol}/L_{Edd}$. Figure 2d (lower right hand corner) is a
scatter plot of the logarithm of $Q$ versus the logarithm of
$L_{bol}/L_{Edd}$.}
\end{figure}
\section{Comparison With Other Results}
Ostensibly, the existence of a high $L_{bol}/L_{Edd}$ and high $Q$
source such as 0743-67 appears at odds with the result of
\citet{wan04}, $Q/L_{bol}$ is inversely correlated with
$L_{bol}/L_{Edd}$. The large Q of Cygnus A appears at odds with
the other conclusion of \cite{wan04}, Q is positively correlated
with $L_{bol}$. However, closer inspection of the raw data used in
\citet{wan04} indicates that this is not actually the case.
\par The virtue of the estimates in \citet{wan04} is that they use the parsec scale jet
emission to estimate $Q$ contemporaneously with the estimate of
$L_{bol}$. However, we warn the reader that such estimates are
very sensitive to the uncertain Doppler factor. The method that
\citet{wan04} adopted from \citet{cel97} assumes that the X-ray
energy emission is from synchrotron self Compton emission (SSC),
however \citet{der93} showed that external Compton (ECS)
scattering of quasar disk photons or broad line region photons by
energetic particles in the jet will usually dominate the high
energy quasar spectrum, since ECS emission is enhanced by the jet
Lorentz factor to the sixth power. This  type of estimator can
lead to enormous errors in the estimated values of $Q$. As an
example, \citet{wan04} estimate for 4C 52.27 (1317+520),
$Q>250Q_{\mathrm{cygA}}$, where $Q_{\mathrm{cygA}}$ is the kinetic
luminosity of Cygnus A. By contrast, using the radio maps in
\citet{hin83} and the isotropic estimator in (3.1), we find a more
reasonable value of $Q\approx 0.35 Q_{\mathrm{cygA}}$.
\par First of all, \citet{wan04} present data in their log-log plot in
figure 1a indicating that $L_{bol}$ and $Q$ have a strong linear
correlation (note that they assume that $L_{bol}\approx 10
L_{BLR}$). However, if one removes the BL-Lacs from the sample and
fit a line to just the quasars on a log-log plot that is otherwise
identical to figure 1a of \cite{wan04} then the squared multiple
regression correlation coefficient, $R^{2}=0.12$. If one removes
the extreme estimate associated 4C 52.27 that was given above, the
linear fit is even worse, $R^{2}=0.04$ and this result is
displayed in figure 2a. This corresponds to a correlation
coefficient, $r=0.2$ and the probability of getting this by chance
is $P=0.174$. The data in figure 2 is lifted directly from
\citet{wan04}, so all the estimates are identical. The data of
\citet{wan04} actually shows that $Q$ and $L_{bol}$ are very
weakly correlated in quasars.
\par The other result of \cite{wan04} that $Q/L_{bol}$ and $L_{bol}/L_{Edd}$
are inversely correlated actually follows trivially as a
consequence of the fact that $Q$ is uncorrelated with $L_{bol}$
for quasars and $L_{bol}/L_{Edd}$ is strongly correlated with
$L_{bol}$. This latter correlation is not surprising; it is the
strongest correlation amongst quasar parameters in \citet{wan04},
the best linear fit is $\log
L_{bol}=0.9309\log[L_{bol}/L_{Edd}]+47.121$ (see figure 2b). The
correlation coefficient is $r=0.820$ and $P<10^{-4}$. Since $Q$ is
uncorrelated with $L_{bol}$, it follows that $Q/L_{bol}\sim\sim
L_{bol}^{-1}$ (where we have introduced the symbol $\sim\sim$ to
represent correlation) and $L_{bol}/L_{Edd}\sim\sim L_{bol}$ from
figure 2b. Combining the two relations, it follows that
$L_{bol}/L_{Edd}\sim\sim L_{bol}/Q$, i.e., $Q/L_{bol}$ and
$L_{bol}/L_{Edd}$ are inversely correlated as shown in figure 2c.
The best linear fit is $\log[Q/
L_{bol}]=-0.814\log[L_{bol}/L_{Edd}]-0.673$ and the correlation
coefficient for $L_{bol}/L_{Edd}$ and $Q/L_{bol}$ is $r=-0.654$
for the subsample of quasars, with $P=3\times 10^{-4}$. The
anti-correlation of $L_{bol}/L_{Edd}$ and $Q/L_{bol}$ is spurious:
there is no direct causal link between the these two variables as
expressed statistically by small value of the partial correlation
coefficient of $Q/L_{bol}$ versus $L_{bol}/L_{Edd}$ with $L_{bol}$
held fixed, -0.030. Finally, we note that this result does not
imply that there is the potentially interesting correlation
between $Q$ and $L_{bol}/L_{Edd}$ as evidenced by figure 2d. The
correlation is very weak, $r=0.1489$ and $P=0.244$.
\section{Conclusion}
PKS~0743$-$67 is an example of a quasar that has an ultra-luminous
accretion flow, $L_{bol}>2\times 10^{47}\mathrm{ergs/s}$, a very
high Eddington rate, $L_{bol}/L_{Edd}\approx 1$ with
$Q>2.5Q_{CygA}$, and is presently active as evidenced by the
powerful unresolved VLBI radio core. By contrast, the high $Q$
source Cygnus A lies at the low end of the quasar range of
$L_{bol}$ and has a small $L_{bol}/L_{Edd}$ and is also presently
active as evidenced by the jet extending from the lobes to within
a few light years of the central black hole (see figure 1.10 of
\citet{pun01}). Using a large sample of quasars in figure 2, it
was shown that $L_{bol}$ is uncorrelated with $Q$. Hence, the
diverse values of $Q/L_{bol}$ in Cygnus A and PKS 0743-67 should
not be unexpected. It appears that to first order, the parameters
$L_{bol}/L_{Edd}$ and $L_{bol}$ are unrelated to the intrinsic
quasar jet power. This is consistent with the observation that
$\approx90\%$ of quasars are radio quiet, from the most luminous
quasars down to the quasar/Sevfert 1 dividing line. Consider the
wide range of $L_{bol}$ in quasars that are associated with very
powerful jets. It is argued in \citet{sem04} and \citet{pun01}
that a significant large scale magnetic flux near a rapidly
spinning black hole is the missing ingredient and is the primary
determinant of FRII quasar jet power, not the accretion flow.

\end{document}